# *Microheater actuators as a versatile platform for strain engineering in 2D materials*

Yu Kyoung Ryu[1], Felix Carrascoso[1], Rubén López-Nebreda[2], Nicolás Agraït[2,3,4], Riccardo Frisenda[1,*] and Andres Castellanos-Gomez[1, *]

[1]Materials Science Factory. Instituto de Ciencia de Materiales de Madrid (ICMM-CSIC), Madrid, Spain.
[2]Departamento de Física de la Materia Condensada, Universidad Autónoma de Madrid, E-28049, Madrid, Spain.
[3]Condensed Matter Physics Center (IFIMAC) and Instituto "Nicolás Cabrera", Universidad Autónoma de Madrid, E-28049, Spain,
[4]Fundación IMDEA Nanociencia, Ciudad Universitaria de Cantoblanco E-28049 Madrid, Spain.

riccardo.frisenda@csic.es , andres.castellanos@csic.es

**Abstract**

We present microfabricated thermal actuators to engineer the biaxial strain in two-dimensional (2D) materials. These actuators are based on microheater circuits patterned onto the surface of a polymer with a high thermal expansion coefficient. By running current through the microheater one can vary the temperature of the polymer and induce a controlled biaxial expansion of its surface. This controlled biaxial expansion can be transduced to biaxial strain to 2D materials, placed onto the polymer surface, which in turn induces a shift of the optical spectrum. Our thermal strain actuators can reach a maximum biaxial strain of 0.64 % and they can be modulated at frequencies up to 8 Hz. The compact geometry of these actuators results in a negligible spatial drift of 0.03 µm/ºC, which facilitates their integration in optical spectroscopy measurements. We illustrate the potential of this strain engineering platform to fabricate a strain-actuated optical modulator with single-layer $MoS_2$.

**Introduction**

Two-dimensional (2D) semiconducting materials can withstand exceptionally large mechanical deformations before breakdown (>10% for transition metal dichalcogenides),[1,2] in contrast to the three-dimensional bulk semiconductors that typically present low failure strains ($\leq 1\%$) due to the presence of lattice and surface defects.[3–5] This mechanical resilience of 2D semiconductors, due to the absence of dangling bonds in their surface, together with a large strain sensitivity of their band structure makes this family of materials of particular interest for strain engineering experiments.[6–13] Interestingly, unlike in strain engineering for bulk semiconductors, which typically relies on applying the strain by forcing the epitaxial growth of materials with dissimilar lattice parameters, for 2D materials there is a rich variety of strategies that allows one to apply a variable strain to them. Moreover, this capability of adjusting the level of strain at will has opened the door to straintronic devices.[14–19]

Here we demonstrate a platform to achieve fast modulation of biaxial strain in atomically thin $MoS_2$. Our approach is based on the fabrication of ring-shaped metallic microheaters on top of polypropylene (PP), a polymer substrate with a very large thermal expansion coefficient. The 2D material to be strained is transferred onto the PP surface, in the middle of the ring. Biasing the microheater we can reliably change the temperature of the PP surface, leading to a controlled biaxial expansion of the PP that is in turn transferred to the $MoS_2$ flakes.[20,21] We found that these thermal micro-actuators respond to frequencies up to 8 Hz, which is a factor of ~100 faster compared to that of macroscopic heaters. Moreover, thanks to their compact design and small heating area, our microheaters present a very low spatial thermal drift of ~0.03 µm/ºC and are reliable in time showing good reproducibility in consecutive thermal cycling measurements. We have also estimated the homogeneity of the strain transferred from the microheater to a $MoS_2$ flake by measuring the spatially resolved differential reflectance across the $MoS_2$ flake. These measurements reveal very homogeneous strain levels across the whole $MoS_2$ flake. We finally exploited the control over the strain to demonstrate a fast and large modulation of the refractive index of single-layer $MoS_2$. All these results point out the superior use of microheaters as a versatile platform to apply and modulate in time biaxial strain in 2D materials in a highly controlled and reversible way.

**Results and discussion**

The ring-shaped metallic microheaters are fabricated on PP substrates by optical lithography, metal deposition and lift-off process. See the Materials and Methods section for the detailed description of the fabrication steps. PP was chosen as substrate because of the combination of its resistance to organic solvents (necessary for the lithographic processing), its high thermal expansion (to yield sizeable biaxial strain upon heating) and its high Young's modulus (to ensure a good strain transfer).[20,22] Figure 1a shows optical images of one of the fabricated microheater thermal actuators.

In the following, we compare the performance of the microheaters with that of macroscopic heaters in terms of thermal drift upon thermal cycling to control biaxial strain in 2D materials. The PP substrate (with the ring-shaped metallic microheater on top) is thermally anchored to a macroscopic Peltier element through thermally conductive tape. Through the text, we denominate '*macroscopic heater configuration*' to the use of a Peltier element to increase the global temperature of the whole substrate. Figure 1b shows the in-plane drift in the X and Y axis upon temperature increase using either the microheater (blue circles) or the macroscopic heater (red circles) configuration. While the observed drift of the microheater is smaller than ~0.03 µm/ºC, the macroscopic heater yields a much sizeable drift of ~0.17 µm/ºC. The top inset in Figure 1b shows the superposition of optical microscopy images acquired at 27 ºC and 80 ºC using the macroscopic heater configuration, where the displacement of the microheater caused by the drift is significant. The bottom inset in Figure 1b shows the superposition of optical microscopy images acquired at $V_{heater} = 0$ V (27 ºC) and $V_{heater} = 1$ V (corresponding to a temperature of 75 ºC in the center of the heater, see Sup. Info. Section S1 for a complete discussion about the calibration of the temperature in its center upon different biasing conditions) using the microheater configuration. In this latest case, the displacement due to the thermal drift is barely noticeable.

The 2D material that will be subjected to biaxial strain can be placed onto the center of the ring-shaped microheaters right after their fabrication by an all-dry deterministic transfer method.[23–26] The inset in Figure 2a shows an optical image in transmission mode of a single-layer $MoS_2$ flake transferred in the inner part of the ring microheater. Figure 2a shows differential reflectance spectra measured on the single-layer $MoS_2$ flake for different bias voltage values applied to the microheater. The differential reflectance spectra display prominent peaks corresponding to the generation of excitons in $MoS_2$.[27,28] We have labeled these features as A and B in accordance with the most extended notation in the literature.[29–31] The red-shift of both A and B excitonic peaks as a function of the applied voltage (-52 meV/V$^2$, corresponding to -1.02 meV/ºC) is a consequence of the temperature increase and the tensile biaxial strain induced on the monolayer by the Joule heating in the microheater.[20,21,32,33] Note that the intrinsic thermal shift of the A and B excitons (without biaxial strain) is -0.4 meV/ºC for single-, bi-, and tri-layer MoS$_2$.[21] Therefore, in Figure 2a a shift of -0.62 meV/ºC can be attributed to the biaxial expansion of the MoS$_2$ lattice.

Figure 2b shows the energy of the excitonic peaks as a function of the as a function of the square of the voltage applied to the heater (as the substrate temperature, which is linearly proportional to the biaxial expansion, is proportional to the power dissipated by Joule heating $V^2/R$). One can convert this voltage squared to actual temperature of the PP surface by comparing the spectral shift displayed in Figure 2b (A and B exciton energy *vs*. $V^2$) with that obtained by turning the microheater OFF and measuring differential reflectance spectra while increasing the temperature using a macroscopic heater equipped with a thermocouple (this measurement yields A and B exciton energy *vs*. temperature of the flake). Dividing the slopes obtained in those two measurements one can get the temperature of the flake as a function of the microheater bias voltage squared (see Section S1 of the Supporting Information for more detailed discussion about the calibration of the flake temperature upon biasing the microheater). The biaxial expansion of the PP substrate (that is transduced to biaxial strain in the flake) can be thus calculated by multiplying the temperature increase (with respect to room temperature) by the PP thermal expansion coefficient of 128·10$^{-6}$ ºC$^{-1}$, determined in previous works.[20,21] Then, we determine the gauge factor of the A and B excitons, that is, the spectral shift per % of biaxial strain, from the slope of the linear fits of the graphs in Figure 2b, after subtracting the intrinsic thermal shift value abovementioned being 48 meV/% and 46 meV/% respectively. Figure S2 of the Supporting Information shows additional measurements performed on a bilayer MoS$_2$ that shows a strain gauge factor of 55 meV/%, 50 meV/% and 60 meV/% for the A, B and IL (interlayer) excitons respectively. Figure S3 in the Supporting Information shows other measurements on a trilayer MoS$_2$ with a strain gauge factor of 32 meV/% and 25 meV/% for the A and B excitons respectively. These values are compatible with those obtained with macroscopic heaters.[20,21] The maximum strain level that can be applied with the microheaters actuators is +0.64% and it is limited by the melting temperature of the PP substrate as at higher microheater biasing conditions the PP substrate close to the metal electrodes starts to melt down.

The reproducibility of the spectral shift induced by the microheater was tested on a trilayer MoS$_2$ flake by applying 30 cycles of heater-OFF ($V_{heater}$ = 0 V) and heater-ON ($V_{heater}$ = 0.52 V) states, which correspond to 0% and 0.21% of biaxial strain respectively (Figure 2c). The A exciton energy reproducibly switches between two well-defined values (~1.88 eV and ~1.86 eV) when the strain is switched between 0% and 0.21% respectively. At the right side of Figure 2c we show a histogram to quantify the reproducibility. Optical images of the device in reflection and transmission optical microscopy modes and the differential reflectance spectra at the two levels of strain can be found in Figure S3 in the Supporting Information.

The uniformity of the induced biaxial strain by the microheater actuator has been checked by spatially mapping the differential reflectance on a bilayer/trilayer MoS$_2$ flake transferred in the center of a microheater. Figure 3a shows the optical image of the flake on the PP surface acquired in reflection mode. A first mapping of the flake, with an area of around 320 µm$^2$, was performed at 0 % strain (room temperature), see the left panel in Figure 3b. The same area was then mapped under a strain of 0.35% (right panel in Figure 3b). The spatial variation of the energy of the A exciton is shown in the colormaps displayed in Figure 3b. Both panels of Figure 3b and the histogram extracted from the maps, shown in Figure 3c, prove the highly uniform strain applied by the thermal actuation of the ring-shaped microheaters. Therefore, although one expects a higher temperature at the close vicinity of the metal electrodes, the temperature in the central part of the microheater does not have a sizeable spatial variation.

Next, we compare the operation speed of the microheater with respect to the macroscopic heater configuration. To do so, we have used the same trilayer MoS$_2$ tested in Figure 2c. We have monitored the position of the A exciton as a function of time while we turn ON and OFF the heaters (Figure 4a). The direct comparison shows that the microheater response is almost instantaneous, being of the same order than the spectra acquisition time of our spectrometer (110 ms), while the macroscopic heater one is around two orders of magnitude slower. This result is another fundamental advantage of microheaters compared to macro-heaters to control the biaxial strain in 2D materials. The response of the microheaters as a function of the frequency was also directly measured by monitoring the modulation amplitude of the A exciton energy while the microheater is biased with a sinusoidal wave of increasing frequency (Figure 4b), obtaining a cutoff frequency of 2.5 Hz and a sizeable response even up to 8 Hz. Similarly, the time resolved response in the resistance of the microheater to a step change in the biasing voltage (from 0.5 V to 0.3 V) shows an exponential decay with a time constant of 0.35 s, that corresponds to a frequency of 2.8 Hz comparable to the cut-off frequency estimated from the frequency response (inset from Figure 4b).

The capability of fast modulating the strain on 2D materials in a reliable way may open possibilities to use these microheater thermal actuators in optical modulation applications.[34–37] Figure 5a shows the time evolution of the differential reflectance spectra of a single-layer $MoS_2$ flake when the microheater is biased with an AC voltage (0.7 $V_{pp}$ and 0.5 Hz). As expected from the previous measurements, the excitonic features present in the differential reflectance spectra are shifted following the AC driving signal fed into the microheater and this modulation is quite reliable over more than 60 cycles. Interestingly, using the Fresnel equations for the three optical media (air/$MoS_2$/PP) combined with a Kramers-Kronig analysis we can extract the two components of the complex refractive index from the measured differential reflectance spectra. Figure 5b and 5c show the time evolution of the real- ($n$) and imaginary-part ($\kappa$) of the complex index of refraction of 1L-$MoS_2$. Both these quantities accurately follow the strain modulation imposed by the microheater thermal actuator. In order to better quantify the magnitude of the strain-induced optical modulation in the single-layer $MoS_2$, Figures 5d, 5e and 5f display linecuts of Figure 5a, 5b and 5c, respectively, at an energy of 1.9 eV (653 nm of wavelength). According to Figure 5d, the AC strain modulation introduces a remarkable variation in the differential reflectance amplitude of 11%, at the probed energy of 1.9 eV. The modulation of $n$ and $\kappa$ also reach 2.5% and 13% respectively, which is significantly larger than the refractive index modulation achieved with thermally-actuated dielectric based optical modulators (1.1%).[37] This motivates the potential use of these microheater thermal strain actuators to fabricate 2D-based optical modulators. The main limitation of these thermal strain actuators, however, is to reach very high operation speed. A simple way to increase the operation speed is to reduce the thickness of the substrate. Indeed the response time of the heater can be expected to be $\sim t^2/\alpha$, where $t$ is the thickness and $\alpha$ is the thermal diffusivity of the PP. A further improvement in the operation speed by a factor of 100-1000 can be achieved by reducing the dimensions of the microheaters. In fact, Romain Quidant and co-workers recently demonstrated that microheaters 10 µm in diameter can operate at frequencies up to ~3.5 kHz.[37]

**Conclusion**

In summary, we introduce the use of thermal actuators based on microheaters fabricated on PP substrate for strain engineering in 2D materials. We demonstrate that these actuators are a versatile and straightforward platform to control the biaxial strain in atomically thin $MoS_2$ layers. We show how these thermal actuators allow to apply a maximum strain of 0.64 % that can be reliably modulated at 8 Hz with a negligible spatial drift of 0.03 µm/ºC. We illustrate the use of these heaters to perform strain modulated reflectance measurements to determine the excitons energies and gauge factors with accuracy and reproducibly in time. We also demonstrate a large strain modulation of the refractive index of single-layer $MoS_2$ using our microheater strain actuator. In conclusion, we introduced a straining platform to apply biaxial strain to 2D materials and nanomaterials in general.

**Materials and methods**

*Microheaters fabrication*: PP substrates of 200 µm thickness (Fellowes, US) have been flattened to ensure the uniformity prior to the lithographic processing. To accomplish this, we fix the PP substrate to a $SiO_2$/Si substrate with kapton tape. Then, we spin-coat TI 35ESX resist (MicroChemicals®, Germany) at 3000 rpm for 30 s on the substrate. The substrate is soft baked at 100ºC for 3 minutes on a hot plate. Then, the substrate is exposed by blue light (430-470 nm) for 2.5-3 s with a 2.5× objective using a Smart Print (Microlight3D, France) maskless photolithography system. After developing the substrates in a solution consisting in 2 parts AZ Developer (MicroChemicals®, Germany): 1 part DI $H_2O$ for 50 s, the exposed parts of the resist are removed. Finally, a 5 nm Ti/ 50 nm Au thick layer is deposited on the substrates by electron-beam evaporation. Finally, the residual photoresist is stripped by lift-off process immersing the samples on a TechniStrip® Micro D350 (Dimethyl sulfoxide, MicroChemicals®, Germany) solution at 60 ºC. For some samples, a gentle sonication for 30 s – 60 s can be applied if required.

*Transfer of the $MoS_2$ few layers to the microheaters*: The $MoS_2$ few-layers were mechanically exfoliated from a natural molybdenite mineral crystal (Molly Hill mine, Quebec, Canada) with Nitto tape (Nitto SPV 224) and then transferred onto Gel-film (Gel-Pak, WF 4× 6.0 mil), which is a commercially available polydimethylsiloxane (PDMS) substrate. Prior to the transfer, the thickness of the flakes was determined by quantitative analysis of transmission mode optical images and by micro differential reflectance spectroscopy.[38,39] After thickness characterization, the chosen flakes were transferred onto the center of the ring-shaped microheaters contained on the PP substrates by dry deterministic transfer method[23–25].

*Differential reflectance measurements*: The optical microscopy images and differential reflection spectroscopy measurements were acquired with a Motic BA MET310-T microscope with a fiber-coupled CCD spectrometer (Thorlabs CCS200/M). The details of the equipment setup are explained elsewhere.[27]

*Spatial mapping of the differential reflectance*: To perform the spatial mapping of the differential reflectance we used a setup similar to the one described above consisting of a Motic BA310Met-H microscope operated in epi-illumination mode with a fiber-coupled CCD spectrometer (Thorlabs CCS200/M, external trigger mode) attached to a TENMA programmable benchtop power supply to supply the trigger voltage. The sample was mounted below the microscope objective onto a X-Y motorized translation stage (EKSMA optics 2x 960-0070-03LS Motorized translation stage and 980-0942 2-axis translational stage controller). We control the TENMA voltage source and the X-Y motorized stage through an home-made routine written in Matlab, which perform a raster scan in the X and Y directions with a user defined step-size and at every point stops to acquire a reflectance spectrum.

*Microheater configuration*: The voltage is applied by a TENMA programmable benchtop power supply and the current vs. voltage (*IV*) curves as a function of the temperature are measured with a source-measure unit (Keithley® 2450).

*Time resolved resistance measurement on microheater:* The time response of the microheater was obtained by measuring the resistance in a 4-terminal configuration using a Zurich Instruments HF2LI lock-in amplifier and an AC voltage at high frequency (~20 kHz) to bias the microheater.

*Macro-heater configuration*: The PP substrate containing the microheater + $MoS_2$ flake are thermally anchored onto a Peltier (TEC1-12706) with thermally conductive tape. This allows us to regulate the temperature of the whole substrate+heater+flake system with the macroscopic heater. The temperature is measured by a thermocouple attached to the Peltier surface in proximity of the sample. We have checked that attaching the thermocouple on the surface of the PP substrate leads to a temperature difference of ~1 ºC with respect to temperature on the Peltier surface.

**Acknowledgements**

This project has received funding from the European Research Council (ERC) under the European Union's Horizon 2020 research and innovation programme (grant agreement n° 755655, ERC-StG 2017 project 2D-TOPSENSE) and the European Union's Horizon 2020 research and innovation program under the Graphene Flagship (grant agreement number 785219, GrapheneCore2 project and grant agreement number 881603, GrapheneCore3 project). R.F. acknowledges the support from the Spanish Ministry of Economy, Industry and Competitiveness through a Juan de la Cierva-formación fellowship 2017 FJCI-2017-32919. NA and RLN acknowledge the funding by Spanish MINECO (MAT2017-88693-R) and Comunidad de Madrid (P2018/NMT-4321).

**Supporting Information**

Supporting Information is available online: calibration of the microheaters power to temperature conversion, additional measurements of biaxial strain in $MoS_2$ flakes, strain-amplified thermoreflectance spectroscopy. This material is available free of charge via the internet at http://pubs.acs.org."

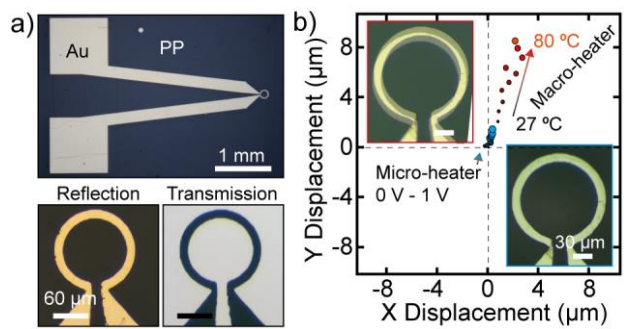

**Figure 1.** (a) Optical images of the fabricated microheater actuators on a PP substrate with different magnifications. (b) Thermal drift induced by ramping up the temperature with a macroscopic heater (red circles) from 27 °C to 80 °C and with the microheater (blue circles) from 27 °C to 75 °C. The top-left inset shows an overlap of two optical images of the sample acquired at 27 °C and 80 °C using the macroscopic heater. The bottom-right inset shows an overlap of two optical images of the sample acquired at 27 °C and 75 °C using the microheater.

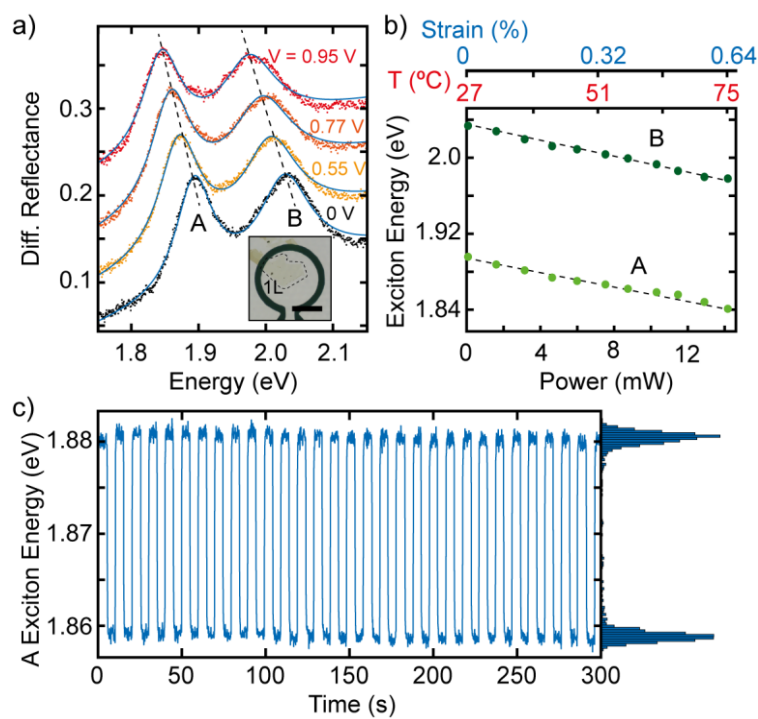

**Figure 2.** (a) Differential reflectance spectra of a single-layer $MoS_2$ flake transferred onto the middle of the microheater actuator as a function of the bias voltage applied to the microheater. (Inset) Transmission mode optical image of the device. (b) Energy of the A and B excitons as a function of the square of the voltage applied to the heater (proportional to the electrical power dissipated in the heater). The top axis shows the equivalent substrate temperature and corresponding biaxial biaxial expansion. (c) A exciton energy as a function of time when the microheater is switched between heater-OFF ($V_{heater}$ = 0 V, $T$ = 27 °C) and heater-ON ($V_{heater}$ =

0.52 V, $T$ = 43 ºC), corresponding to biaxial strain switched between 0% and 0.21%, using a square signal. The histogram at the right side illustrates the reproducibility of the switching.

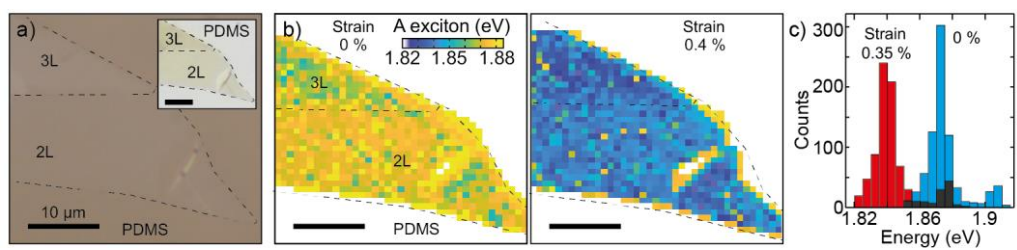

**Figure 3.** (a) Reflection mode optical image of a bilayer/trilayer MoS$_2$ flake transferred onto the middle of a microheater actuator. The inset shows the transmission mode optical image of the same flake before transferring it to facilitate the identification of the different regions. (b) Spatial map of the A exciton energy acquired at two different strain levels (0% and 0.35%). (c) Histogram of the A exciton energy values that show very well-defined peaks, illustrating the spatial homogeneity of the induced strain.

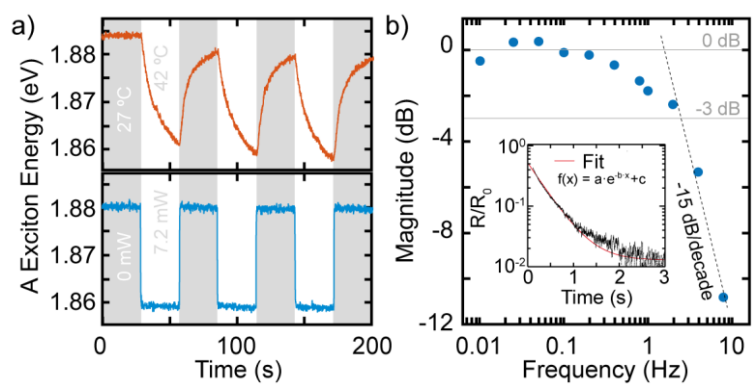

**Figure 4.** (a) A exciton energy as a function of time while the heater is switched ON and OFF. In the top panel the results of using a macroscopic heater are shown and in the bottom panel we show the same experiment switching ON and OFF a microheater thermal actuator. (b) Frequency response of the microheater actuator. The inset shows the normalized resistance change of a microheater when the bias voltage is suddenly reduced from 0.5 V to 0.3 V. The data have been fitted to an exponential decay function (red curve).

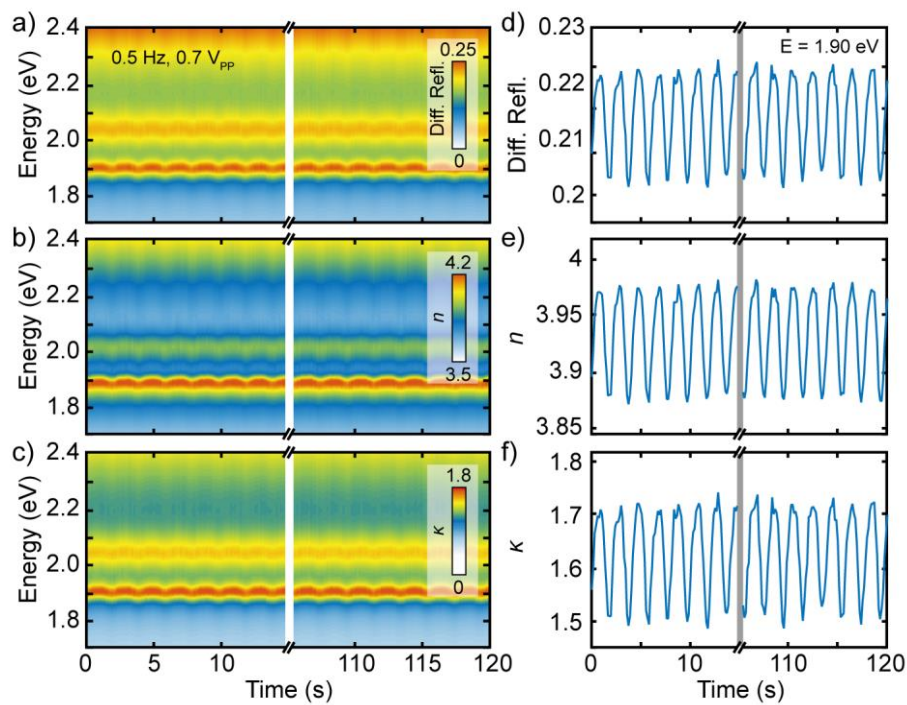

**Figure 5.** (a) Colormap of the differential reflectance of a single-layer MoS$_2$ flake applying a sine wave modulated voltage to the microheater (frequency 0.5 Hz, amplitude 0.7 V$_{pp}$, sampling rate 10 Hz). (b-c) Colormaps of the real part *n* (b) and imaginary part *κ* (c) of the complex refractive index of single-layer MoS$_2$ calculated from the differential reflectance map of panel (a). (d-f) Horizontal linecuts extracted from the maps of panels (a-c) at an energy of 1.9 eV.

# *Supporting Information: Microheater actuators as a versatile platform for strain engineering in 2D materials*

Yu Kyoung Ryu[1], Felix Carrascoso[1], Rubén López-Nebreda[2], Nicolás Agraït[2,3,4], Riccardo Frisenda[1,*] and Andres Castellanos-Gomez[1, *]

[1]Materials Science Factory. Instituto de Ciencia de Materiales de Madrid (ICMM-CSIC), Madrid, Spain.
[2]Departamento de Física de la Materia Condensada, Universidad Autónoma de Madrid, E-28049, Madrid, Spain.
[3]Condensed Matter Physics Center (IFIMAC) and Instituto "Nicolás Cabrera", Universidad Autónoma de Madrid, E-28049, Spain,
[4]Fundación IMDEA Nanociencia, Ciudad Universitaria de Cantoblanco E-28049 Madrid, Spain.

riccardo.frisenda@csic.es , andres.castellanos@csic.es

**Summary**

Section S1- Calibration of the microheaters power to temperature conversion

Section S2- Additional measurements of biaxial strain in MoS$_2$ flakes

Section S3- Strain-amplified thermoreflectance spectroscopy

**Section S1- Calibration of the microheaters power to temperature conversion**

To calibrate the microheaters described in the main text we performed a set of measurement of the transport characteristics of the microheater shown in Figure 2a, both as a function of the bias voltage and as a function of temperature. We also recorded the differential reflectance spectrum of the single-layer MoS$_2$ transferred in the middle of the microheater ring as a function of the bias voltage and temperature. Figure S1a shows a set of *I-V*s of the microheater circuit recorded at different global temperatures between 28 ºC and 68 ºC (the temperature was controlled by a Peltier element and monitored through a thermocouple). The voltage range has been chosen small on purpose to avoid self-heating of the wire. The *I-V*s appear linear and the resistance, given by the inverse of the slope, increases while increasing the temperature, as expected for a normal metal. Figure S1b shows the resistance of the microheater as a function of the temperature together with a linear fit, which gives a temperature coefficient of 0.135 Ω/ºC.

If one increases the voltage beyond a certain point, self-heating effects become important and the microheater can dissipate power into heat and warm up its surroundings. Figure S1c shows an *I-V* characteristic recorded at room temperature (T = 28 ºC) while slowly increasing the voltage from 0 V to 0.65 V (at a rate of 0.005 V/s). While at low bias the *I-V* appears to follow a linear trend, at larger voltages it starts to deviate from this trend. Figure S1d shows the resistance versus voltage calculated from the *I-V* of the previous panel, by dividing at each point the voltage for the current. The resistance clearly shows a parabolic behavior increasing together with the voltage, a sign of the self-heating process, as confirmed also by the good agreement between the data and the fit to a second order polynomial. We find a voltage coefficient of 7.11 Ω/V$^2$.

The differential reflectance spectrum of the single-layer MoS$_2$ flake transferred in the middle of the ring, as already discussed in the main text, shows peaks coming from direct excitonic resonances that are closely related to the MoS$_2$ bandgap energy. Figures S1e shows the differential reflectance spectra of the single-layer MoS$_2$ flake recorded at room temperature (blue curve) and at global temperature of 88 ºC (red curve), controlled with a macroscopic heater equipped with a thermocouple. The figures also show multi-Gaussian fit to the data (three Gaussian peaks + linear background) used to extract the energy of the excitons. Figure S1f shows similar measurements performed at room temperature respectively with the heater kept at 0 V (blue curve) and applying a voltage of 0.95 V (red curve). Both figures show a clear red-shift of the peaks either when increasing the temperature of the macroscopic heater or when increasing the bias voltage of the microheater. We quantify this red-shift by extracting the A and B excitons position as a function of macroscopic heater temperature or microheater bias voltage and then performing a linear fit, as shown in Figures S1g and S1h. Notice that in Figure S1h we plotted the square of the voltage in order to linearize the data (as the temperature increase induced by the microheater is expected to be proportional to the Joule heating power $V^2/R$). From the linear fit we find respectively an exciton red-shift of -1.14 meV/ºC and -56.0 meV/V$^2$. The results of these and of the previous fits are summarized in Table S1. From the four experimental quantities just discussed we can extract two values of the temperature-voltage coefficient of our microheater, that ideally should coincide. From the microheater transport measurements we find 52.7 ºC/V$^2$ and from the MoS$_2$ optical measurements we find 49.1 ºC/V$^2$, given the good agreement between the two estimations we choose to take the average of the two values for the calibration (50.9 ± 1.8) ºC/V$^2$.

|  | Microheater transport | MoS$_2$ excitons |
|---|---|---|
| Temperature (Peltier) | 0.135 Ω/ºC | -1.14 meV/ºC |
| Voltage | 7.11 Ω/V$^2$ | -56.0 meV/V$^2$ |

**Table S1.** Fit parameters extracted from the measurements reported in Figure S1.

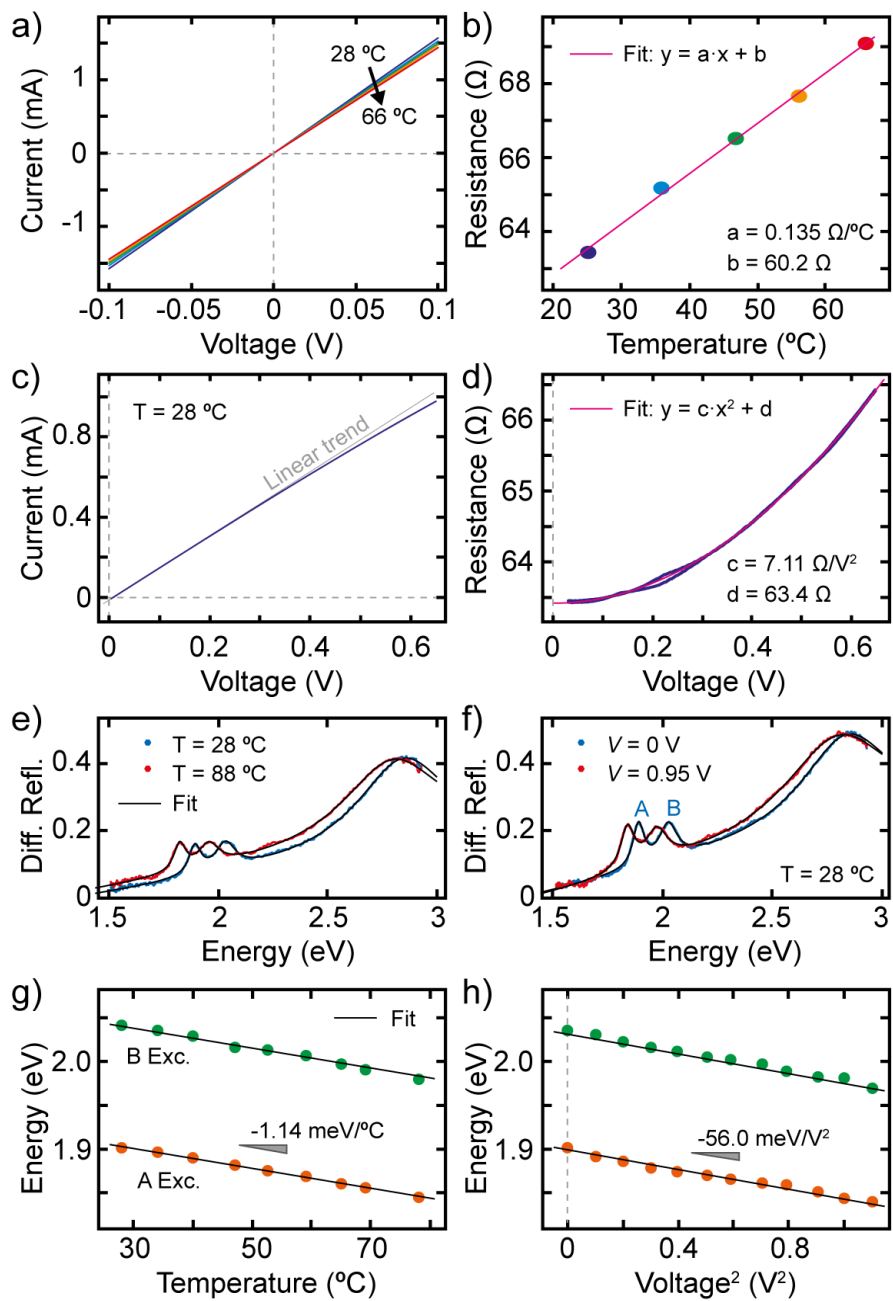

**Figure S1.** (a) Current-voltage characteristics (*I-V*s) of the microheater circuit recorded as a function of global temperature (controlled with a macroscopic heater). (b) Resistance extracted from the *I-V*s in panel (a) as a function of temperature. The magenta line is a linear fit to the data. (c) *I-V* characteristic of the microheater recorded at room temperature while sweeping the voltage from 0 V to 0.65 V at a rate of 0.005 V/s, the grey line is a guideline for the eye representing a linear trend. (d) Resistance of the microheater as a function of voltage, the magenta line is a fit to a second order polynomial. (e-f) Differential reflectance spectra of a single-layer MoS$_2$ transferred in the middle of the ring at two different global temperatures (e) and two different applied voltages on the microheater (f). The black lines represent multi-Gaussian peaks fits. (g-h) Energy of the A and B excitons as a function of global temperature (e) and voltage squared (f) with linear fits to the data.

**Section S2- Additional measurements of biaxial strain in MoS₂ flakes**

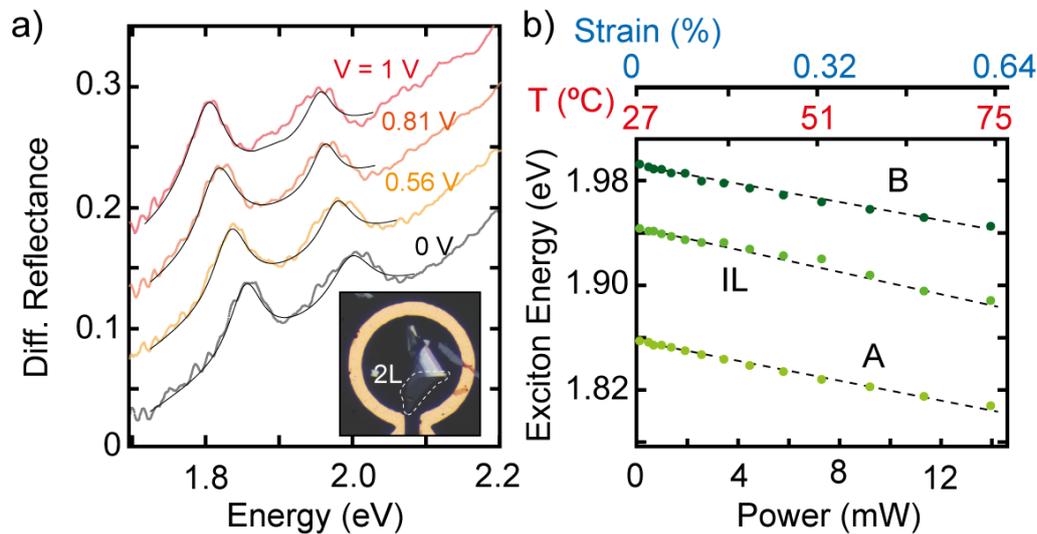

**Figure S2.** (a) Differential reflectance spectra of a bilayer MoS$_2$ transferred in the middle of the ring of a microheater recorded with increasing applied voltages on the microheater. (b) energy of the A, B excitons as a function of the power dissipated in the microheater.

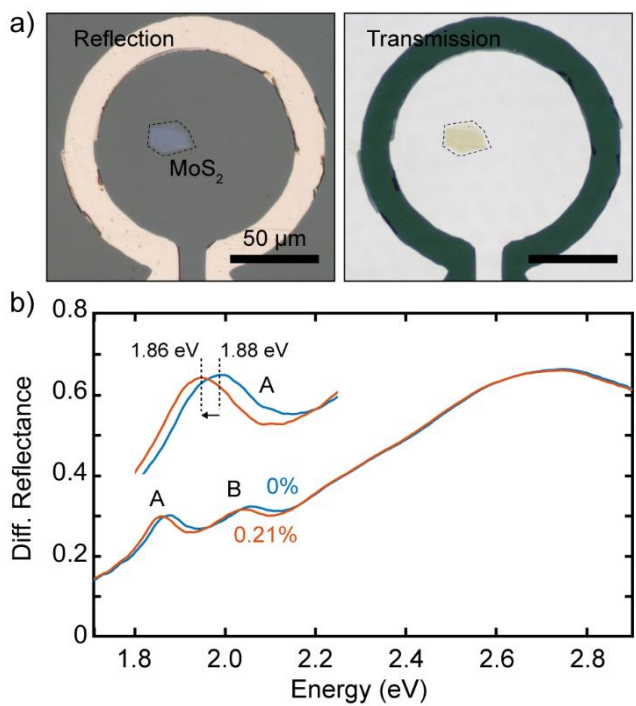

**Figure S3.** (a) Reflection (left) and transmission (right) mode optical image of a MoS$_2$ flake transferred onto the middle of a microheater actuator. (b) Differential reflectance spectra of the trilayer MoS$_2$ region at 0% and 0.21% of strain.

**Section S3- Strain-amplified thermoreflectance spectroscopy**

In this section we illustrate the potential use of the microheater actuators for fast modulation of the biaxial strain in 2D materials in time. Figure S4a shows the time evolution of the differential reflectance spectra of the monolayer MoS$_2$ shown in Figure 2a by applying an AC voltage (at a frequency of 0.5 Hz) to the microheater: 0.3 V$_{pp}$ (left panel) and 0.7 V$_{pp}$ (middle panel). The right panel of Figure S4a displays the intensity of the differential reflectance spectra, extracted at 1.87 eV (the point with highest slope in the A exciton peak) and 1.89 eV (the center of the A exciton peak), as a function of time for the case of $V_{heater}$ = 0.3 V$_{pp}$ (left panel). From these traces, one can already see that the sinusoidal modulation induced in the reflectance depends on the energy, in our case 1.87 eV shows a large and periodic modulation of the amplitude while at 1.89 eV we only observe a constant value within the experimental resolution.

Interestingly, one can obtain information about the strain-tunable features of the spectra by using the fast Fourier transform (FFT) analysis of the reflectance spectrum intensity *vs.* time datasets at different energies. Figure S4b shows how the vertex of the excitonic peaks show a negligible FFT magnitude at the modulation frequency while the spot with larger slope of the peak show strong peaks in the FFT at the modulation frequency and at its first harmonic (also visible in the two time-traces displayed in the right panel in Figure S4a). This can be very useful for certain optical spectroscopy experiments. Pinpointing one of these possible applications, we demonstrate how a fast-sinusoidal strain modulation can be used to accurately determine the energy of the different excitonic features in the MoS$_2$ reflectance spectra even if there is a large background noise. In a similar fashion to other modulation spectroscopy techniques such as piezo reflectance or electro reflectance,[40] we apply a sinusoidal signal to the microheater that yields to a sinusoidal variation of the biaxial strain, leading to a time varying spectral shift. If we repeat the FFT analysis shown in Figure S4b for each energy value, we can construct a spectrum of the strain-varying spectral features. Particularly, in this representation the vertex of the excitonic peaks show up in this technique as strong dips (Figure S4c top) while the spots with high slope in the differential reflectance show up as strong peaks in this representation. This allows to determine the A and B exciton energies with accuracy. Figure S4c compares the spectrum obtained by modulating strain (top) with that obtained by conventional differential reflectance (bottom). One can clearly see how the strain modulation allows for an accurate determination of the excitonic vertex even for very broad excitonic features with a large background signal. Note that the presence of an extra dip between those related to the A and B excitons is expected when two strain-tunable peaks are close to each other. Figure S5 in the Supporting Information shows a simulation of the FFT analysis in presence of noise and for more than one peak in the differential reflectance.

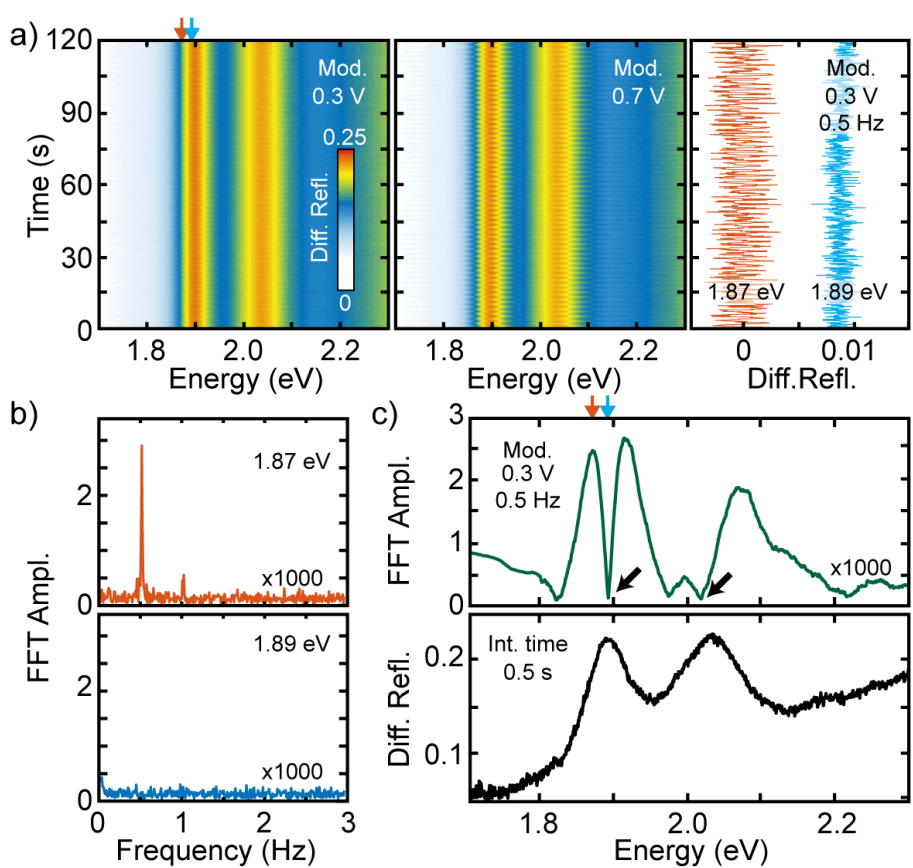

**Figure S4.** (a) Differential reflectance spectra as a function of time when the strain is modulated with a sine wave at 0.5 Hz: left panel $V_{mod} = 0.3$ $V_{pp}$, middle panel $V_{mod} = 0.7$ $V_{pp}$. The right panel shows the differential reflectance intensity vs. time extracted from the left panel at 1.87 eV and 1.89 eV (indicated by the red and blue arrows). (b) Fourier transform spectra of the two signals shown in the right panel in (a). (c, top) Spectrum constructed with the FFT amplitude at 0.5 Hz for different energies. This spectrum provides information about the strain-tunable features present in the differential reflectance spectrum (e.g. the vertex of the strain-tunable exciton peaks should show up as strong dips, indicated with black arrows). (c, bottom) Conventional differential reflectance spectrum for comparison.

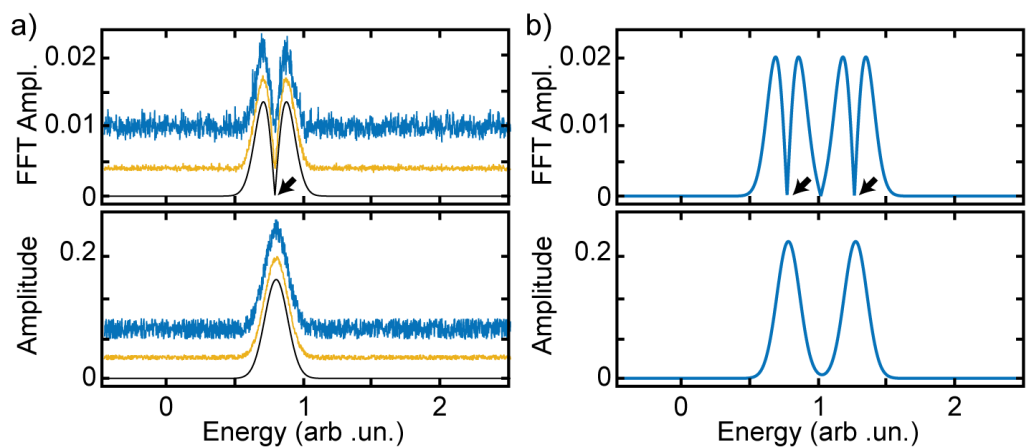

**Figure S5.** (a) Amplitude of the FFT transform of the fundamental frequency (top) and original signal (bottom) consisting of a single Gaussian peak with additional levels of white noise whose center is modulated in time by a sine wave. (b) Same as (a) for two close Gaussian peaks without noise.